\documentstyle[sprocl]{article}

\input{psfig}
\def\be{\begin{equation}}
\def\ee{\end{equation}}
\def\bea{\begin{eqnarray}}
\def\eea{\end{eqnarray}}

\newcommand{\nc}{\newcommand}
\nc{\renc}{\renewcommand}
\nc{\np}[3]{{\em  Nucl.\ Phys.\ }{{\bf #1} {(#2)} {#3}}}
\nc{\pr}[3]{{\em Phys.\ Rev.\ }{{\bf #1} {(#2)} {#3}}}
\nc{\prl}[3]{{\em Phys.\ Rev.\ Lett.\ }{{\bf #1} {(#2)} {#3}}}
\nc{\pl}[3]{{\em  Phys.\ Lett.\ }{{\bf #1} {(#2)} {#3}}}
\nc{\prep}[3]{{\em Phys\. Rep.\ }{{\bf #1} {(#2)} {#3}}}
\def\lsim{\; \raise0.3ex\hbox{$<$\kern-0.75em
      \raise-1.1ex\hbox{$\sim$}}\; }
\def\gsim{\; \raise0.3ex\hbox{$>$\kern-0.75em
      \raise-1.1ex\hbox{$\sim$}}\; }
\def\vec#1{{\bf #1}}
\nc{\cL}{{\cal L}}
\nc{\nn}{\nonumber \\*}
\def\GeV{{\rm\ GeV}}
\nc{\G}{\rm \ G}
\newcommand{\rf}[1]{(\ref{#1})}
\begin{document}

\title{MAGNETIC FIELDS IN THE EARLY UNIVERSE\footnote{Invited Talk at the 
Strong and Electroweak Matter '97
21-25 May 1997, Eger, Hungary}}

\author{KARI ENQVIST}

\address{Department of Physics, P.O. Box 9, FIN-00014 University
of Helsinki, Finland}


\maketitle
\abstracts{The observed galactic magnetic fields may have a primordial
origin. I briefly review the observations, their interpretation in terms of 
the dynamo theory, and the current limits on cosmological magnetic fields.
Several possible mechanisms for 
generating a primordial magnetic field are then discussed. Turbulence and the 
evolution of the microscopic fields to macroscopic  fields is described
in terms of a shell model, which provides an approximation
to the full magnetohydrodynamics and indicates the existence of an
inverse cascade of magnetic energy. Cosmological seed fields 
roughly of the order of $10^{-20}$ G at the scale of protogalaxy, 
as required by the dynamo explanation
of galactic magnetic fields, seem rather plausible.}
\section{Introduction}
Apart from the baryon number and the spectrum of energy density fluctuations,
the physical processes that took place in the very early universe do not 
have many consequences that could  still be directly detectable today.
Most observables have been washed away by the thermal bath of the
pre-recombination era. One possibility, which has recently received
increased attention, is offered by the large-scale magnetic fields 
observed in a number of galaxies, in galaxy halos, and in clusters of 
galaxies  \cite{observe}. 
The astrophysical mechanism
responsible for the origin of the galactic magnetic fields is not 
understood. Instead one postulates a small seed field, which can then be either
enhanced by the compression of the protogalaxy, 
or exponentially amplified by the
turbulent fluid motion as in the dynamo theory  \cite{dynamo}. The
exciting possibility is that the seed field could be truly primordial,
in which case cosmic magnetic fields could provide direct information
about the very early universe.

The problem then is twofold. First, one has to find a mechanism in the early 
universe which is able to produce a magnetic field large enough to act
as the seed field. There are various proposals, a number of which are
based on the early cosmological phase transitions, which are discussed
in Sect 3. The second problem is to explain how the initial field, 
which is expected to be random as it is created 
by microphysics and having correlation lengths typical to microphysics,
can grow up to be coherent enought at large length scales. This is a
problem in magnetohydrodynamics which is discussed in Sect 4.
\section{Observations and limits}
\subsection{Observing cosmic magnetic fields}
Magnetic fields at the level of few $\mu{\rm G}$
have been detected in galaxies, in galactic halos, and
in clusters of galaxies.
They can be observed
indirectly at optical and radio wavelengths \cite{observe}. The 
Zeeman splitting of the spectral lines would provide a direct
measure of the strength of the magnetic field, but the shifts
are very small and this method is 
applicable mainly to our own galaxy. 

Electrons moving in a magnetic field emit synchrotron radiation,
and both its intensity and polarization can be used to extract
information about the magnetic field. However, one needs first to fix
 the relative magnitudes of the electron and
magnetic field densities. Usually equipartition of 
magnetic and plasma energies is assumed.

Information about distant magnetic fields in e.g. clusters in
galaxies has been obtained by studying the Faraday rotation of polarized
light.
The method is based on the fact that
the plane of polarization of linearly polarized electromagnetic wave
rotates as it passes through plasma supporting a magnetic field. The
rotation angle $\Delta \chi$ depends on the strength and extension of the magnetic field,
the density of plasma, and on the wavelength $\lambda$ of radiation.
The Faraday rotation measure (RM) is defined as
\be
{\Delta \chi \over \Delta \lambda^2} = 8.1\times 10^5\int n_eB_{l}dl
~{\rm rad\;m^{-2}}~,
\ee
where $B_{l}$ is the strength of the magnetic field along the line of sight.
This method requires some independent information about the electron
density and the field reversal scale.

\subsection{$\alpha-\omega$ dynamo}
The currently favoured explanation for the origin of the large scale 
galactic magnetic fields is the $\alpha-\omega$ dynamo \cite{dynamo},
which through turbulence and differential rotation amplifies a small
frozen-in
seed field ${\bf B}_0$ to the observed $\mu{\rm G}$ field. 
An initially toroidal 
seed field, which is carried along on the disc of a rotating (spiral) galaxy,
is locally distorted into a loop by the up- or downward stochastic drift of the
gas. As the gas moves away from the plane of the disc, the pressure
decreases and the gas expands; at the same time it is subject to a Coriolis force which will rotate it. The magnetic field lines, glued to the gas,
will follow and thus a poloidal component perpendicular to ${\bf B}_0$
is generated. The small poloidal loops will reconnect and coalesce
to produce a large scale field. This is
the so-called $\alpha$-effect. Because the disc does not rotate like a 
rigid body, the field lines will be wrapped and the poloidal field
will induce a toroidal field; this is the $\omega$-effect. 
Thus the dynamo mechanism does not only produce a large scale
field but can also, to some extent, predict the shape of the field.
In practise one does not attempt to describe the small scale turbulence
directly but rather resorts to a mean-field dynamo, where the average
induced current is given by ${\bf j}_{ind}=\sigma\alpha{\bf B}_0$,
where $\sigma$ is conductivity and $\alpha$ is a parameter
related to the average drift velocity.

The dynamo saturates when the growth enters the non-linear regime.
A typical growth time is of the order of $10^9$ years, with the
rotation period of the galaxy setting a lower limit on the growth
time. Recently it has been argued \cite{dproblems}
that the saturation might actually 
be too fast for a large-scale field to form.
The strength of the required initial seed field is rather
uncertain, but as a rule of thumb one could use a value like
$10^{-20}$ G on a comoving scale of a protogalaxy (100 kpc).

Another possibility is that the galactic field results
directly from a primordial field, which gets compressed when
the protogalactic cloud collapses. The primordial field strengths
needed are however quite large.
In any case, it appears as if a primordial field is required to
explain the observed galactic magnetic fields 
(although it is conceivable that a purely astrophysical solution could 
exist as well). 
\subsection{Limits on cosmologial fields}
Observing cosmological, intergalactic magnetic fields would be of great
importance and would strengthen the case for their truly primordial
origin. It has been claimed by Plaga \cite{plaga}
that the arrival times of extragalactic
$\gamma$-rays could be used to detect fields as weak as $10^{-24}$ G.
The idea is that 
cosmic rays, originating from far-away objects such as
QSOs or gamma-ray bursters, 
scatter  off the background cosmic magnetic field. This gives rise to pair
production and a delayed $\gamma$-ray which could then be observed, and
the ratio of prompt to delayed $\gamma$'s provides a measure of the
strength of the intergalactic magnetic field. The energy 
spectrum of $\gamma$-rays
is likewise affected by the intergalactic magnetic field \cite{los}.
Ultra-high cosmic ray protons 
would also be deflected by the intergalactic magnetic
fields so that their arrival times could be used \cite{cray} 
to set bounds as low
as $10^{-12}\G$.

A direct limit on cosmological magnetic field
has been obtained by considering the rotation measure of a sample of 
galaxies and quasars, which yields the limit \cite{vallee} $B \le 10^{-9} \G
\; (\Omega_{IG}h_{100}/0.01)^{-1}$, where $\Omega_{IG}$ is the fraction of
the ionized gas density of the critical density in the intergalactic medium,
and $h$ is the Hubble constant. It has also been suggested that 
the power spectrum of the cosmological
field could be determined by studying the correlations in the RM
for extragalactic sources  \cite{kolatt}.

If a primordial magnetic field is present at the time of recombination,
it will give rise to anisotropic pressures. These would distort the
microwave background. Barrow, Ferreira and Silk \cite{cmblimit}
have recently considered the effect of a magnetic field on the evolution
of shear anisotropy in a general anistropic flat
universe. They then compared the result with the 4-year COBE data
set, and assuming that the whole observed anisotropy is due to
magnetic stresses, concluded that 
if the field can be taken homogeneous, 
$B \le 3.4\times10^{-9} \G
\; (\Omega_{0}h_{50}^2)^{1/2}$. 
This is a more stringent limit
than what can be obtained  \cite{nuclsynt} from primordial nucleosynthesis
considerations. These take into account 
the contribution of the magnetic field to the Hubble expansions rate
and to the $n\leftrightarrow p$ reaction rates, plus the fact
that the electron wave functions get modified (the phase space
of the lowest Landau level is enhanced). The limit thus obtained is
$B\le 10^{11}\G$ for a constant $B$
at $T=10^9$ K, and 
if the field is inhomogeneous, the limit is less stringent by an 
order of magnitude.

The anisotropy limit also indicates
that the  distortions of the Doppler peaks of the microwave 
background \cite{adamsetal} due to a homogeneous magnetic field 
are unobservable. The Faraday rotation in the  polarization
of the microwave background could still be observable \cite{kosowsky}.

\section{Generating primordial magnetic fields}
There are a number of proposals for the origin of  magnetic fields
in the early universe, many of them involving the early phase transitions.
Fluctuations in the electromagnetic field in a relativistic plasma
are by themselves sufficient for generating a small scale random magnetic
field \cite{lemoine}. To obtain a seed field for galactic magnetic fields,
one however needs fields with much larger coherence lengths.
\subsection{Spontaneous charge separation}
Magnetic fields may arise in a electrically neutral plasma if
local charge separation happens to take place, thus creating a local 
current. It has been proposed that this could occur
during a first order QCD \cite{qcdsep} or EW \cite{ewsep,siglolinto}
phase transition, which proceed by nucleating bubbles of
the new phase in the background of the old phase. 
There one finds net baryon number gradients 
at the phase boundaries, providing the basis for charge separation, 
and the seed fields arise through 
instabilities in the fluid flow. For that one has to require
that the growing modes are not damped.
Turbulent flow near the walls of the bubbles is then expected to amplify
and freeze the transient seed field. 
The various hydrodynamical
features have been studied 
in linear perturbation theory by Sigl, Olinto and Jedamzik \cite{siglolinto}, 
who argued that 
on a 10 Mpc comoving scale, field strengths of the order of
$10^{-29}$ G for EW and $10^{-20}$ G for QCD could be obtained.
These might further be enhanced by several orders of magnitude 
by hydromagnetic turbulence \cite{beo1,beo2},
which will be discussed in Section 4.3. 
\subsection{Bubble collisions}
In a first order phase transition the phases of the complex order parameter
$\Phi = \rho e^{i\Theta}/\sqrt{2}$ of the nucleated bubbles are 
uncorrelated. When
the bubbles collide, there arises a phase gradient which acts as
a source for gauge fields. The phase itself is not a gauge invariant
concept, but it has been pointed out by Kibble and Vilenkin \cite{kv}
that a gauge invariant phase difference 
can be defined in terms of an
integral over the gradient $D_\mu\Theta$.

Magnetic field generation in the collision of phase transition bubbles
has been considered in the abelian Higgs model \cite{kv,ae3}. One  assumes that 
inside the bubble the radial mode
$\rho$ settles rapidly to its equilibrium value $\eta$ and can thus be treated
as a constant. The dynamical variables are thus $\Theta$ and the gauge
field $A^\mu$.
The starting point is the U(1)-symmetric lagrangian
\be
\cL=-\frac{1}{4}F_{\mu\nu}F^{\mu\nu}+D_{\mu}\Phi (D^{\mu}\Phi )^{\dagger}+
V(\vert\Phi\vert ),
\ee
where the potential
$V$ is assumed to have minima at $\rho=0$ and $\rho=\eta$.
The simplest case is that two spherical bubbles nucleate, 
one at $(x,y,z,t)=(0,0,z_0,0)$ and the other at $(x,y,z,t)=(0,0,-z_0,t_{0})$,
and keep expanding with the velocity $v$ even after colliding. Because of the 
symmetry of the problem, the solutions to the equations of motion are
functions of $z$ and $\tau=\sqrt{t^2-x^2-y^2}$ only, and the U(1) gauge field
$A^{\mu}=x^{\mu}f(\tau ,z)$. With the appropriate
initial conditions, the solutions are \cite{kv}
\bea
\Theta &=& \frac{\Theta_{0}R}{\pi\tau}\int_{-\infty}^{\infty}\frac{dk}{k}
\sin {k(z-z_1)}
\big[\cos \omega (\tau -R)+\frac{1}{\omega R}\sin \omega (\tau -R)\big]~,\nn
f&=&\frac{\Theta_{0}Re\eta^2}{\pi\tau^3}\int_{-\infty}^{\infty}
\frac{dk}{k}\sin {k(z-z_1)}
\big[\frac{R-\tau}{\omega^2 R}\cos\omega (\tau -R) \cr
&+&\big(\frac{\tau}{\omega}+
\frac{1}{\omega^3 R}\big)
\sin \omega (\tau -R)\big],
\eea
where $\omega\equiv\sqrt{k^2 + e^2\eta^2}$, $R$ is the 
radius of the bubbles at the collision and $z_1$ the point of first collision
on the z-axis. Here the velocity of the bubbles has been taken to be $v=1$.
The generated magnetic field is rapidly oscillating and orthogonal to
the z-axis and in this case confined inside the intersection region,
as can be seen from Fig. 1. It has a ring-like shape in the
(x,y)-plane.
It can be shown that subsequent collisions of the bubbles, which now may 
have a magnetic field inside the bubbles,  nevertheless lead to
a qualitatively similar outcome  \cite{ae3}. 

\begin{figure}
\leavevmode
\centering
\vspace*{70mm} 
\includegraphics{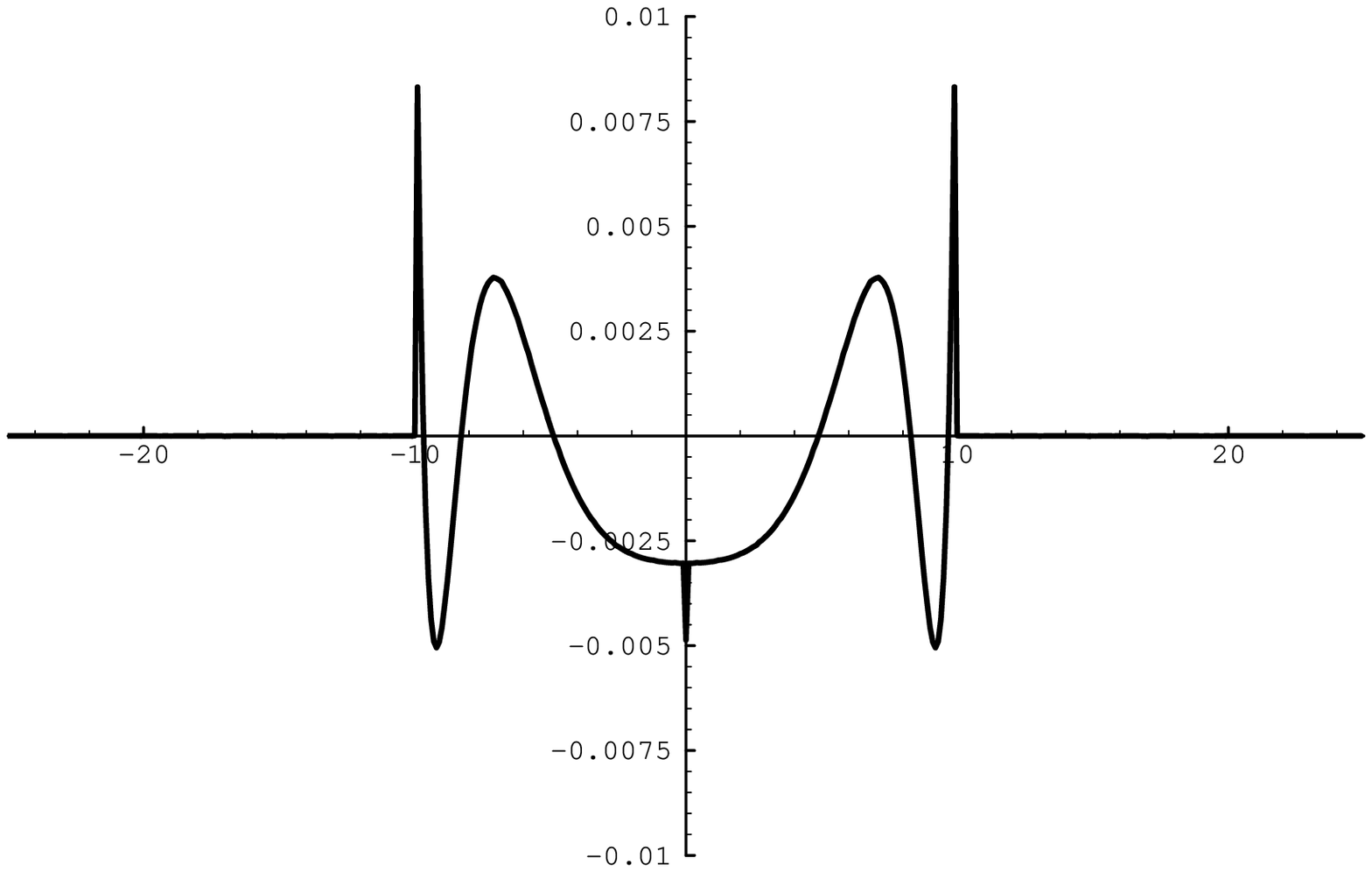}
\includegraphics{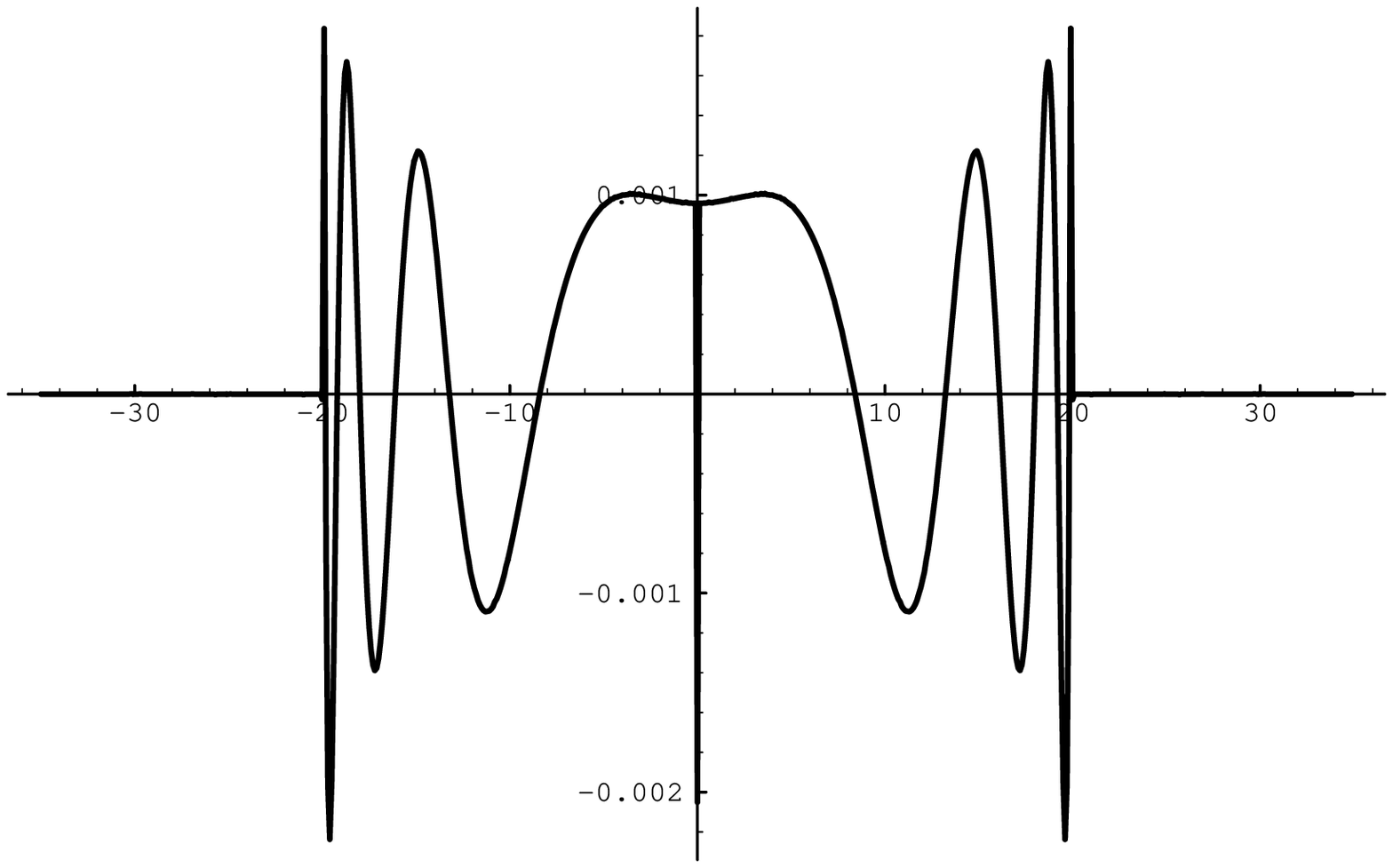}   
\includegraphics{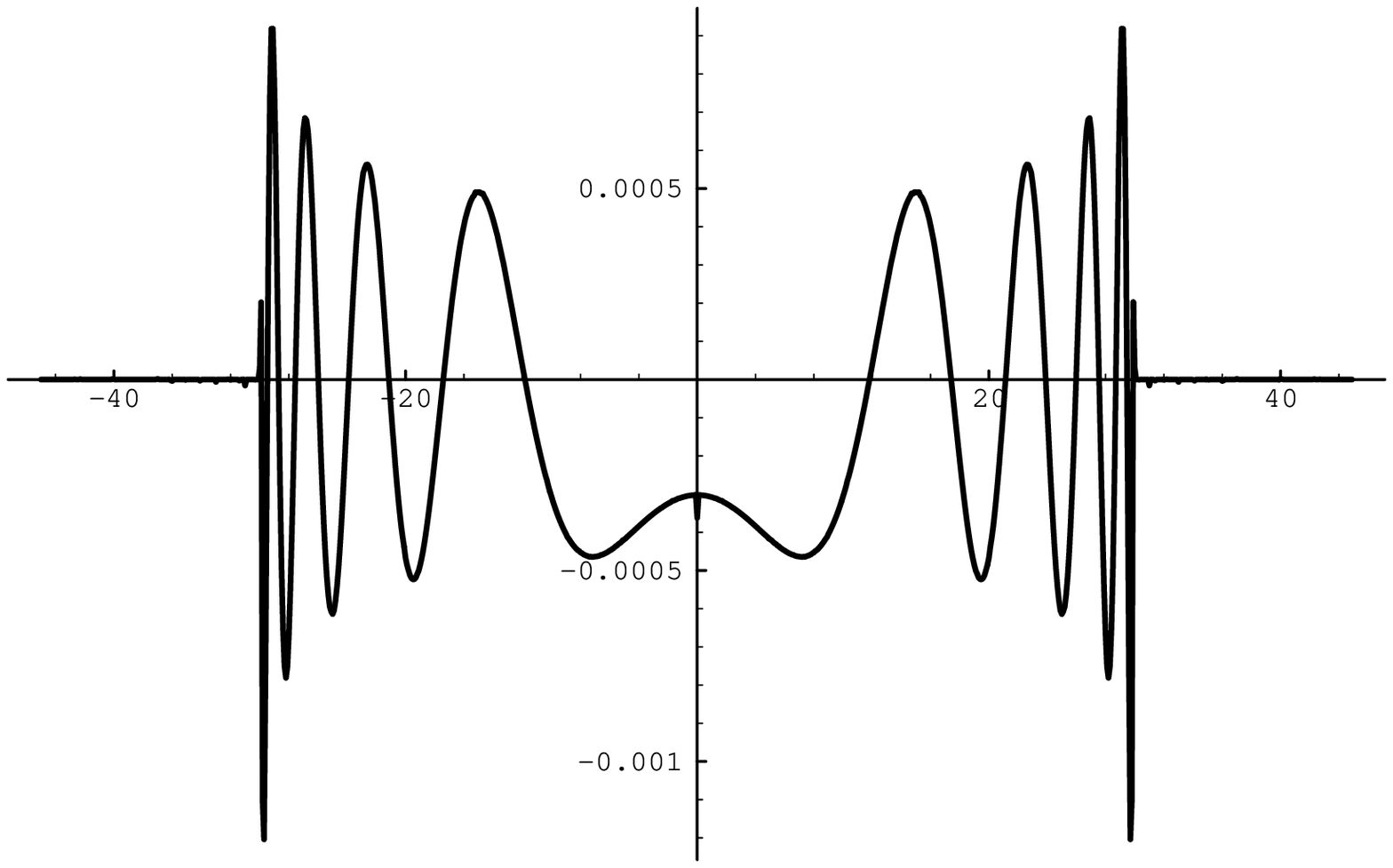}
\includegraphics{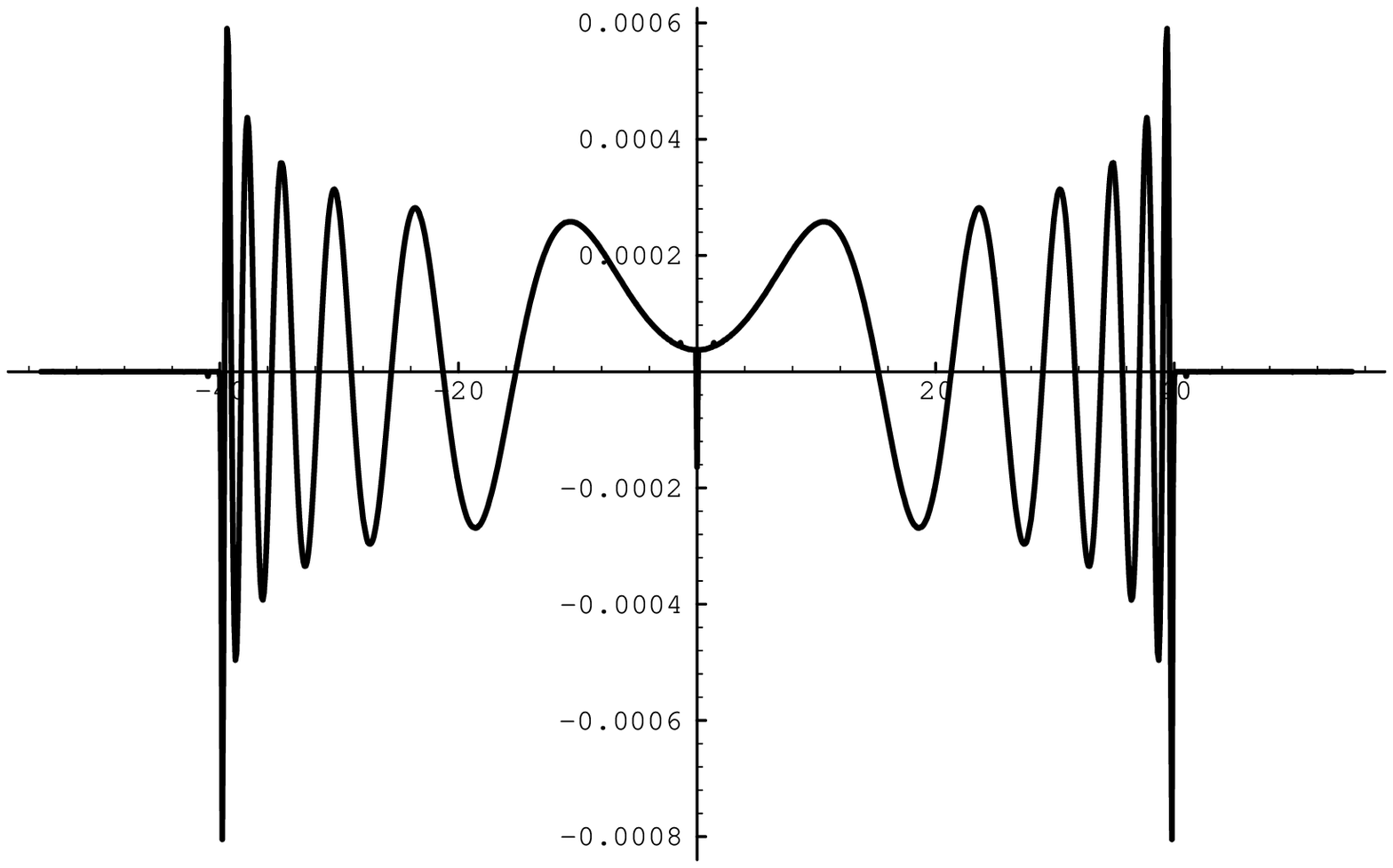}
\caption{The time evolution of 
$B$ along the radius $r=1$ for $t$=10, 20, 30 and 40 after the initial
collision. The radius of the bubbles at the collision has been chosen here 
$R=10$, and the collision point on the z-axis is $z_1=50$.
(The  units are such that $e\eta =1$).}
\label{jkuva2}       
\end{figure} 

There are however two additional important ingredients which need to be
taken into account: the high but finite conductivity \cite{turnerwidrow,cond}
of the primordial plasma
  and the fact that in the electroweak
phase transition the bubbles will in fact intersect with non-relativistic
velocities  \cite{velocity}. Finite conductivity gives rise to diffusion,
the consequence of which
is to smooth out the rapid oscillations
of $\vec B$, whereas low $v$ will permit the magnetic flux to escape
the intersection region and penetrate the colliding bubbles, where its
evolution will
be governed by usual magnetohydrodynamics. The resulting 
behaviour is demonstrated
in Fig 2.

\begin{figure}
\leavevmode
\centering
\vspace*{45mm}
\includegraphics{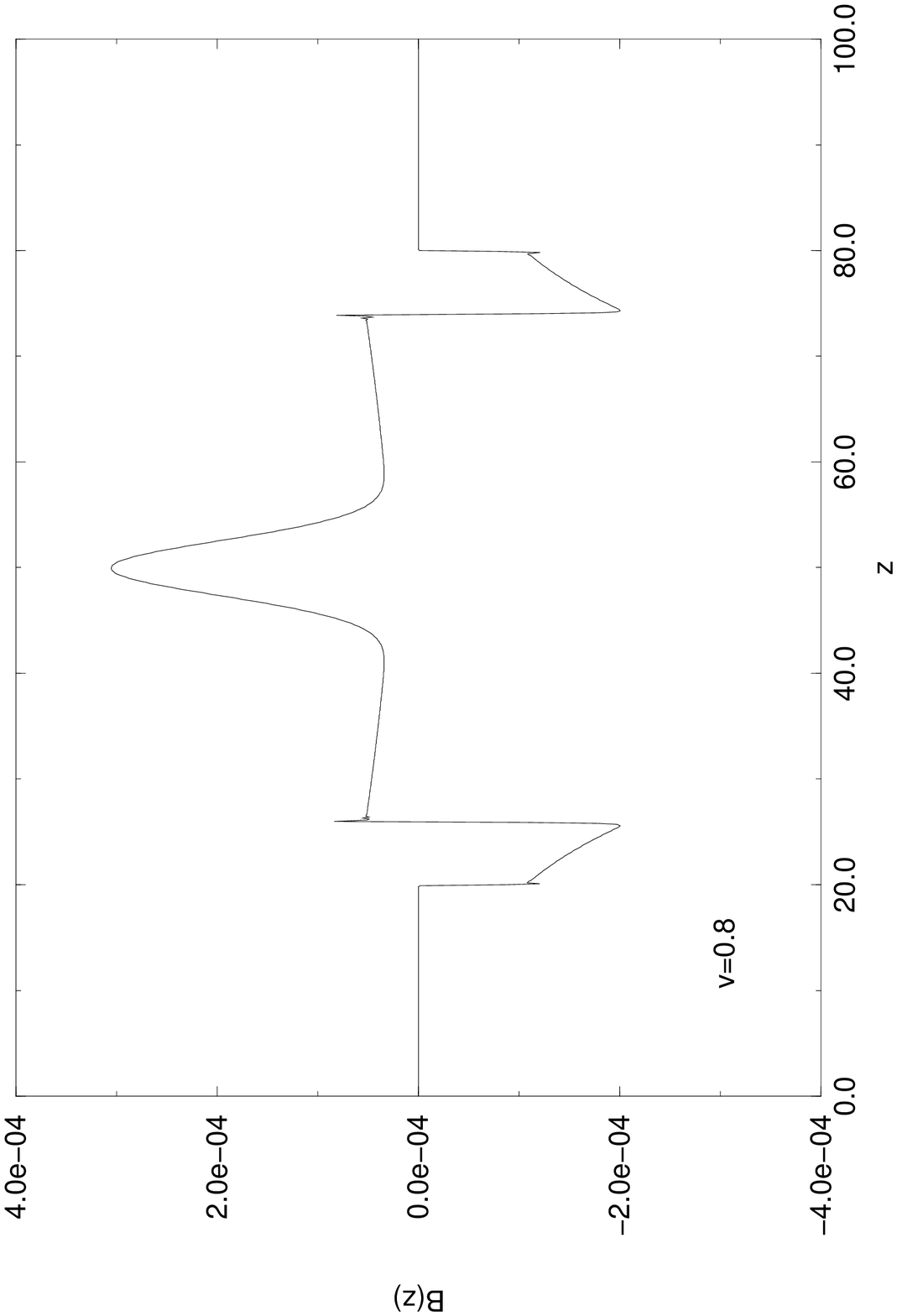}
\includegraphics{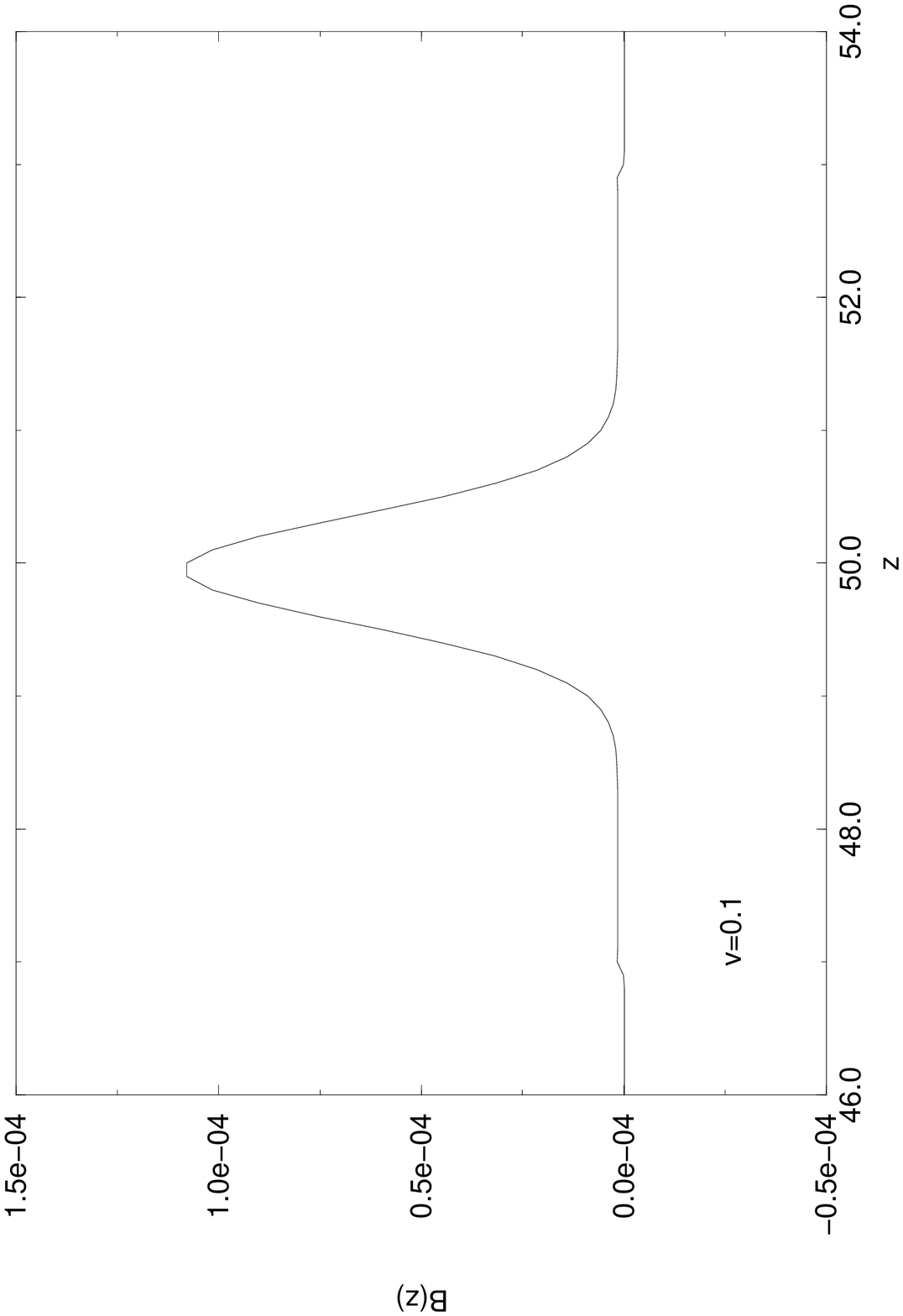}
\caption{$B$ along $r=1$ with $R=10$ at
$t=30$ after collision with
$v=0.8$ and $0.1$. Here conductivity has been taken to be $\sigma =7T$
 The point of initial collision on the 
z-axis is $z_1=50$, and the outer edge of the intersection region is at 
$\simeq 50\pm vt$. The outer edge of the magnetic field is at $\simeq 50\pm t$ 
(and units are $e\eta =1$).}  
\label{kuva8}
\end{figure}

The strength of the generated magnetic field depends on the bubble
wall velocity in an essential way  \cite{ae3}.
In the electroweak case the initial growth of the
bubble wall is by subsonic deflagration, with velocities of the order 
of 0.05$c$,
depending on the assumed friction strength
  \cite{velocity}. The wall is preceded by a shock
front, which may collide with the other bubbles. This results in reheating, and
oscillations of the bubble radii, but
eventually a phase equilibrium is attained. The ensuing bubble growth is
very slow and takes place because of the expansion of the universe.
Because the universe has been reheated back to $T_c$,
no new bubbles are formed during the slow growth phase. 
Assuming that the abelian Higgs model results are applicable to the 
electroweak case, as seems to be the case \cite{dariotoni},
one may estimate that \cite{ae3}
\be
B \simeq 2.0\times 10^{20}
\sqrt{\gamma^2+2\gamma R}/R \, 
\G, 
\ee
where it has been assumed $\Theta_0=1$, $T_c=e\eta=100\GeV$;
the average distance between the nucleation 
centers 
$r_{ave}=9.5\times 10^{-8}t_H$ and the velocity $v=1.2\times 10^{-4}$
were taken as reference values.
Folding in the spectrum of separation of the adjacent shocked spherical 
bubbles  \cite{meyer}, averaging over all possible inclinations of
the ring-like magnetic field, and taking into account the
enhancement of magnetic energy due to an inverse cascade (see next Section),
one arrives at the estimate $B_{rms}\simeq 10^{-21} {\rm G}$ for 
the cosmological magnetic field at the scale of 10 Mpc today \cite{ae3}.
\subsection{Fluctuating Higgs gradients}
It has been pointed out by Vachaspati \cite{vach} that fluctuating Higgs field
gradients will induce a magnetic field. Such local fluctuations
are naturally present at the EW phase transition, since the embedding of
the electromagnetic field in  
$SU(2) \otimes U(1)_Y$ involves these gradients:
\bea
F_{ij}^{em} & = & - i(V^\dagger_i V_j - V_i V^\dagger_j)~~, \nn
V_i & = & \frac{2}{|\phi|} \sqrt{\frac{\sin \theta}{g}}~
\partial_i \phi ~~,                                           
\eea
where $\phi$ is the Higgs field. (An argument against horizon-size
fluctuations has been presented by Davidson \cite{davidson}).
At the electroweak phase transition the correlation length in the
broken phase is
$\sim 1/m_W$ (assuming that the Higgs mass is comparable to
$m_W$).
The field strength $F_{ij}^{em}$ is thus constant over a distance 
$\sim 1/m_W$, but
it varies in a random way over larger distances in order to respect
causality.
The vector $V_i$ is also random, of course.
Its variation is due to the fact that the Higgs field $\phi$ makes a
random walk on the vacuum manifold of $\phi$.
The problem then is to estimate the field strength 
$F_{ij}^{em}$ over a length scale
$\sim N/m_W$.
If $N = 1$, then it follows on dimensional grounds 
that the root-mean-square $F_{ij}^{em} \sim
m_W^2 \sim 10^{24} $ G.
For $N$ large, one should use a statistical argument.
Vachaspati argued that the gradients are of order
$1/ \sqrt{N}$, since $\phi$ makes a random walk on the vacuum manifold
with $\Delta \phi \sim \sqrt{N}$, and since $\Delta x \sim N$.
Thus $V_i$ is, in a root mean square sense, of the order
$1/ \sqrt{N}$, and hence $F_{ij}^{em}$ is of order $1/N$.
Taking further into account that the flux in a co-moving circular
contour is constant, the field must decrease like $1/ a(t)^2$, where
$a(t)$ is the scale factor. However,
there  also exists a different statistical scenario, based on a line
average,
where the gradient vectors are taken to be the basic stochastic 
variables \cite{poul}.
The field strength
can be interpreted statistically in such a way that the
mean magnetic field satisfies
\be
\langle F_{ij}^{em}\rangle_T ~= 0~~,~~~~~\sqrt{\langle F^2_{ij}\rangle_T}~ \sim 
\frac{T^2}{\sqrt{N}}~~.                         \label{4}
\ee
One observes that the scaling behavior is weaker by a factor $\sqrt{N}$.
This means that for a scale of 100 kpc
\be
\sqrt{\langle F^2_{ij}\rangle_{today}} \sim  10^{-18} {\rm G}~~,   \label{5}
\ee
which is very close to the value desired for the dynamo
effect. Because the Faraday rotation involves the average along the
line of sight, this averaging method could indeed be appropriate for those
observations.
\subsection{Vacuum condensates}
Another, more exotic possibility for generating primordial magnetic fields is
based on the observation that, due to quantum fluctuations,
the Yang--Mills vacuum is unstable in a large enough background magnetic
field \cite{savvidy} at zero temperature. There are indications
from lattice calculations that
this is a non-perturbative result \cite{tw}. In a pure SU(N) theory
at the one--loop level the zero temperature  effective energy for a constant
background non--abelian magnetic field reads \cite{savvidy}
\be
V(B)=\frac 12 B^2+ \frac{11N}{96\pi^2}g^2B^2\left(\ln {gB\over\mu^2}-
\frac 12\right)              \label{1}
\ee
with a minimum at
\be
gB_{\rm min}=\mu^2\exp \left(-{48\pi^2\over 11Ng^2}\right)  \label{2}
\ee
and $V_{\rm min}\equiv V(B_{\rm min})=-0.029(gB_{\rm min})^2$.
Thus the ground state
has a non--zero  non--abelian magnetic field, the magnitude
of which is exponentially suppressed relative to the renormalization
scale, or the typical momentum scale of the system. In the early
universe, however, where possibly a grand unified symmetry is
valid, the exponential suppression may be less severe. It is also
attenuated by the running of the coupling constant.

In the early universe the effective energy picks up thermal corrections
from the fermion, gauge boson, and Higgs boson loops. The detailed form of
the thermal correction depends on the actual model, but one may take the
cue from the SU(2) one--loop calculation, where they are
obtained by summing the Boltzmann factors $\exp(-\beta E_n)$ for the
oscillator modes
\be
E_n^2=p^2+2gB(n+\frac 12)+2gBS_3+m^2(T), \label{21}
\ee
where $S_3=\pm 1/2~(\pm 1)$ for fermions (vectors bosons).
Eq. \rf{21} includes the thermally induced mass $m(T)\sim gT$,
 corresponding to
a ring summation of the relevant diagrams. Numerically, the effect of
the thermal mass turns out to be very important.
At high temperature, the leading behaviour is
given mainly by the bosonic contributions, and thus one arrives at 
the estimate \cite{ferro}
\be
\delta V_T^v=\frac {(gB)^2}{8\pi^2}\sum_{l=1}^{\infty}\int_0^{\infty}
\frac{dx}{x^3} e^{-K_l^b(x)}\left[\;
x{{\rm cosh}(2x)\over {\rm sinh}(x)}-1\right],  \label{4}
\ee
where $K_l^a(x)={gBl^2/(4xT^2)}+{m_a^2x/(gB)}$ and numerically 
$\delta V_T^v\sim 0.02\;(gB)^2$ which serves only to shift the value of
$B$ at the vacuum slightly. Thus $B\ne 0$ is the state of the lowest energy
even at high temperature.

A local fluctuation will then trigger the creation of the
a new vacuum with non-zero non--abelian magnetic field 
inside a given particle horizon at scales $\mu \simeq T$.
 The Maxwell magnetic field
is then just a projection of the non-abelian field,
the order of magnitude of which is given by Eq. \rf{2}. 
One can then estimate that \cite{ferro}
\be
B(T)=g_{\rm GUT}^{-1}\mu^2\exp \left(-{48\pi^2\over 11Ng^2}\right)
\left({T^2\over \mu^2}\right)\simeq
3\times 10^{42}\left({a(t_{\rm GUT})\over a(t)}\right)^2\;\G, \label{5}
\ee
where  the reference
number is for susy SU(5). 

Electroweak magnetic condensates have also  recently been considered by 
Cornwall \cite{Cornwall}, who has suggested that they would give rise
to magnetic fields with a net helicity
via the generation of electroweak Chern-Simons number. 
Joyce and Shaposnikov \cite{joyceshapo}
have argued that a
right-handed electron asymmetry, generated at the GUT scale,
could give induce a hypercharge magnetic field via
a Chern-Simons term. Both suggestions deserve further study.
\subsection{Inflation and magnetic fields}
The basic problem with inflation with regards to magnetic field generation 
is that the early universe was a good conductor so that, ignoring turbulence, 
the magnetic flux $\sim Ba^2$ 
tends to be conserved, where $a$ is the cosmic scale factor. 
To avoid this, one needs to break the conformal
invariance somehow, as was first suggested by Turner and Widrow
\cite{turnerwidrow}, who considered couplings to the
curvature $R$ such as $RF^2$ and
$RA^2$, as well as photon-axion couplings. 
Dilaton coupling
of the form $e^\Phi F^2$ has also been considered \cite{ratra}, and interesting
field strengths can be obtained at the expense of tuning the 
coupling strength. If a phase transition takes place during the
inflationary period, a sufficiently large magnetic field can be created,
provided however that the phase transition takes place during the final 
5 e-foldings \cite{davisdimoinfl}.
\section{From microscopic to macroscopic}
Even assuming that a primordial magnetic field is created at some very
early epoch, a number of issues remain to be worked out before
one can say anything definite about the role of primordial fields
for the origin of galactic magnetic fields. At the 
earliest times magnetic fields
are generated by particle physics processes with length scales typical to
particle physics. (It has been shown that such fields are stable against
thermal fluctuations \cite{martindavis}.)
The remaining question is whether it is at all possible
for the small scale fluctuations to grow to large scales, and what exactly is
the scaling behaviour of $B_{\rm rms}$ or the correlator
$\langle B(r+x) B(x)\rangle$.  To
study these problems one needs to consider the detailed evolution of the
magnetic field  to account for such issues as to what happens when uncorrelated
field regions come into contact with each other during the course of the
expansion of the universe. In general, turbulence is an essential
feature of such phenomena. These questions can only be answered by considering
magnetohydrodynamics (MHD) in an expanding universe  \cite{relMHD}.
\subsection{MHD in curved space}
Let us consider the early universe as consisting of ideal fluid with
an equation of state of the form $p=\frac{1}{3}\rho$, where $p$ is pressure
and $\rho$ the energy density. Let us further
assume that the fluid supports a (random) magnetic field.
The energy-momentum tensor is then given by
\begin{eqnarray}
T^{\mu\nu}&=&(p+\rho) U^\mu U^\nu + p g^{\mu\nu} \nonumber \\
&+&{1\over4\pi}\left(F^{\mu\sigma} {F^\nu}_\sigma
-{1\over4}g^{\mu\nu}F_{\lambda\sigma}F^{\lambda\sigma}\right),
\label{energymomentum}
\end{eqnarray}
and $U^\mu$ is the four-velocity of the plasma,
normalized as $U^\mu U_\mu=-1$,  and $F_{\mu\nu}=
\partial_\mu A_\nu-\partial_\nu A_\mu$ is the electromagnetic field tensor.
Note that, as long as diffusion can be neglected, the presence of the magnetic
field does not change the equation of state. 
The magnetic energy is further assumed to be much smaller than the radiation
energy, so that one can  assume a flat, isotropic and homogeneous 
universe with a Robertson-Walker metric.
 Although the magnetic field generates local
bulk motion, this may still be consistent with isotropy and homogeneity at
sufficiently large scales, in particular if the magnetic field is random,
i.e.\ statistically homogeneous and isotropic on scales much larger than the
intrinsic correlation scale of the field.

For numerical treatment 
 it is convenient to write the equations of motion explicitly in 
3+1 dimensions and use the conformal time $\tilde{t}\equiv\int dt$/$a$
as a variable. 
After a lengthy derivation one arrives at the forms \cite{beo1}
\begin{equation}
{\partial\tilde{{\bf S}}\over\partial \tilde{t}}=
-(\mbox{\boldmath $\nabla$}\cdot{\bf v})\tilde{{\bf S}}
-({\bf v}\cdot\mbox{\boldmath $\nabla$})\tilde{{\bf S}}
-\mbox{\boldmath $\nabla$}\tilde{p}+\tilde{\bf J}\times\tilde{{\bf B}}.
\label{tld2}
\end{equation}
and 
\begin{eqnarray}
{2\gamma^2+1\over 4\gamma^2(2\gamma^2-1)}{\partial\ln\tilde{\rho}\over
\partial\tilde{t}}
&=&-{\partial\tilde{{\bf S}}^2/\partial\tilde{t}\over\left({4\over3}\tilde{\rho}
\gamma\right)^2 (2\gamma^2-1)} \nonumber \\
&-&{\bf v}\cdot\mbox{\boldmath $\nabla$}
\ln(\tilde{\rho}\gamma^2)-\mbox{\boldmath $\nabla$}\cdot{\bf v}
+{\tilde{\bf J}\cdot\tilde{\bf E}\over{4\over3}\tilde{\rho}\gamma^2},
\label{rhoexpression2}
\end{eqnarray}
where we have set $\rho+p=\frac{4}{3}\rho$.
The Maxwell equations can be written explicitly as
\begin{equation}
{\partial \tilde{{\bf B}}\over\partial\tilde{t}}=
-\nabla\times\tilde{\bf E},\quad\mbox{\boldmath $\nabla$}\cdot\tilde{{\bf B}}=0,
\label{induct1}
\end{equation}
and
\begin{equation}
\tilde{\bf J}=\mbox{\boldmath $\nabla$}\times\tilde{{\bf B}}
-{\partial \tilde{\bf E}\over\partial \tilde{t}},
\quad\mbox{\boldmath $\nabla$}\cdot\tilde{\bf E}=\tilde\rho_e
\end{equation}
where $\rho_e$ is the charge density and $\tilde\rho_e=a^3\rho_e$. Further,
\begin{equation}
\tilde{\bf E}=-{\bf v}\times\tilde{{\bf B}},
\label{vxB}
\end{equation}
which is valid in the limit of high conductivity.

The evolution of a random magnetic field configuration in 2d is shown 
in Fig. 3 for a lower and higher resolution. As time goes on, the coalescence 
of magnetic structures is seen to lead to the
gradual formation of larger and larger scales. Such a behaviour is encouraging,
indicating that the field is not really comovingly frozen.
On the other hand, 2d MHD is very different from the 3d MHD, and moreover,
the Reynolds number in the simulation (about 10) is wildly unrealistic.
\begin{figure}[htbp]
\psfig{figure=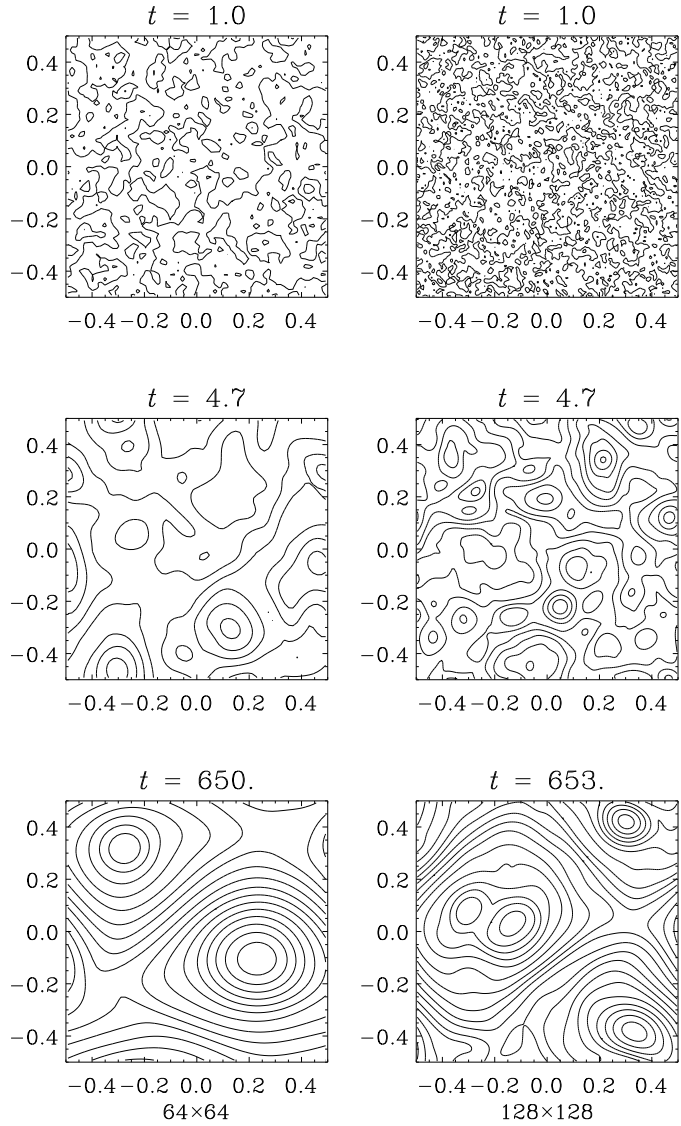,height=6in}
\caption{
Left column: magnetic field lines at different times at low resolution
($64\times64$ meshpoints).
Right column: magnetic field lines at different times at higher resolution
($128\times128$ meshpoints).
}\label{ppsnap}\end{figure}
However,
Dimopoulos and Davis \cite{dimodavis} have also pointed out that when two initially uncorrelated
domains come into contact, the field at the interface should untangle 
with the plasma bulk velocity $v$ to
avoid the creation of domain walls. They propose that the correlation
length  $\xi$ evolves according to
\be
{d\xi\over dt}=H\xi + v~,\label{41}
\ee
where $H$ is the Hubble parameter and 
the velocity $v$ depends dynamically on $B$ and should,
in principle, be determined from MHD. Nevertheless, \rf{41} again 
points towards the possibility of the magnetic field not
necessarily  being comovingly frozen.
\subsection{Shell models}
In ordinary hydrodynamics many properties of
turbulence, in particular those related  to energy transfer and to the spectral
properties have been studied
successfully using a simple cascade model.
This is true not only qualitatively, but also quantitatively, which is
the reason why the cascade model is now much used in studies of
nonlinear physics \cite{mogensbook}.

The basic idea  is that the interactions due to the nonlinear terms
in the MHD equations
are local in wavenumber space, and in $k$-space the quadratic nonlinear terms
become a convolution.
Interactions in $k$-space involving triangles with similar
side lengths have the largest contribution.
This has led to the shell model
which is formulated in
the space of the modulus of the wave numbers. This space is approximated by
N shells, where each shell consists of wave numbers with $2^n\leq k \leq
2^{n+1}$ (in the appropriate units). The Fourier transform of the velocity
over a length scale $k_n^{-1}$ ($k_n=2^n$) is given by the complex quantity
$v_n$, and $B_n$ denotes a similar quantity for the $B$-field.
Furthermore, the convolution is approximated by a sum over the nearest and
the next nearest neighbours,
\begin{equation}
N_n(v,B)=\sum_{i,j=-2}^2 C_{ij} v_{n+i} B_{n+j}.
\end{equation}
Here $v$ and $B$ have lost their vectorial character, which
reflects the fact that this model is not supposed to be an approximation
of the original equations, but should be considered as a toy model that
has similar {\it conservation} properties  as the original equations.

Velocity and magnetic fields are thus represented by scalars at the discrete
wave numbers $k_n=2^n$ ($n=1,...,N$), i.e. $k_n$ increases exponentially.
Therefore such a model can cover a large range of length scales
(typically up to ten orders of magnitude).
The important conserved quantity is $E_{\rm tot} R^4$, where
$E_{\rm tot}=\int T^{00}d^3x$ is the total energy. The 
resulting equations of motion
read \cite{beo1}
\begin{equation}
{\textstyle{4\over3}}\rho_0
{dv_n\over d\tilde{t}}=N_n(v,b),
\label{cascadeu}
\end{equation}
\begin{equation}
{db_n\over d\tilde{t}}=M_n(v,b),
\label{cascadeb}
\end{equation}
where
\begin{equation}
\begin{array}{lll}
2N_n(v,b)&=ik_n(A+C)
(v^*_{n+1}v^*_{n+2}-b^*_{n+1}b^*_{n+2})\\
                        &\!\!\!+ik_n(B-{\textstyle{1\over2}}C)
(v^*_{n-1}v^*_{n+1}-b^*_{n-1}b^*_{n+1})\\
                        &\!\!\!\!\!\!\!-ik_n({\textstyle{1\over2}}B+
{\textstyle{1\over4}}A)
(v^*_{n-2}v^*_{n-1}-b^*_{n-2}b^*_{n-1}),
\end{array}
\end{equation}
\begin{equation}
\begin{array}{lll}
M_n(v,b)&=ik_n(A-C)
(v^*_{n+1}b^*_{n+2}-b^*_{n+1}v^*_{n+2})\\
                        &\!\!\!+ik_n(B+{\textstyle{1\over2}}C)
(v^*_{n-1}b^*_{n+1}-b^*_{n-1}v^*_{n+1})\\
                        &\!\!\!\!\!\!\!-ik_n({\textstyle{1\over2}}B
                                            -{\textstyle{1\over4}}A)
(v^*_{n-2}b^*_{n-1}-b^*_{n-2}v^*_{n-1}),
\end{array}
\end{equation}
with $A$, $B$, and $C$ being free parameters.
It is straightforward to verify that $2\sum v^*_n N_n + \sum b^*_n M_n=0$,
using that $k_n=2^n$. 
\subsection{Inverse cascade}
The numerical study of the cascade model requires of course that the
parameters $A,B,C$ are fixed so that the model has the same conservation
laws as the full-fledged MHD. The model can then be solved 
numerically \cite{beo1}, and
the results
are shown in
Fig. 4, where the 
 transfer
of magnetic energy to larger and larger length scales is clearly seen. 
This process, the inverse cascade, is due to the
nonlinear terms giving rise to mode interactions. The initial magnetic
energy spectrum was chosen to be given by $E_M(t=0)\sim k$ with the 
total magnetic energy equal to $\rho_0$. The number of shells was $N=30$ 
so that length scales differing by ten orders of magnitude were covered.
\begin{figure}
\leavevmode
\centering
\vspace*{105mm}
\includegraphics{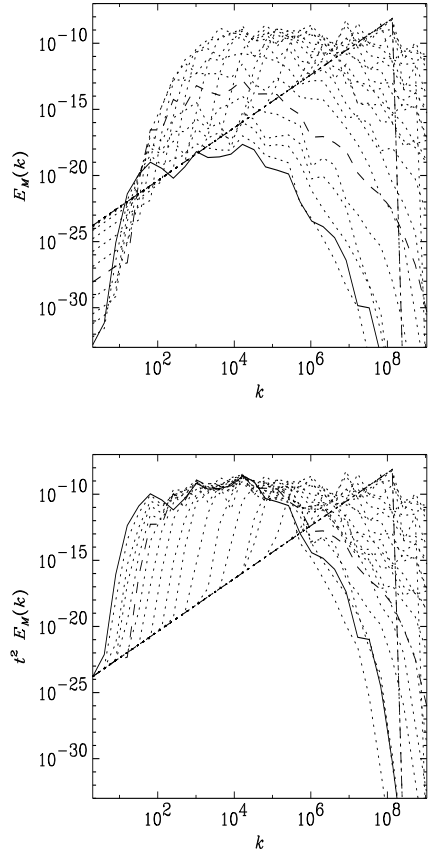}
\caption{
Spectra of the magnetic energy at different times.
The straight dotted-dashed line gives the initial condition ($t_0=1$),
the solid line gives the final time ($t=3\times10^4$), and the dotted
curves are for intermediate times (in uniform intervals of 
$\Delta\log(t-t_0)=0.6)$. $A=1$, $B=-1/2$, and $C=0$.
}\label{pb_array}\end{figure}
It was
found that the integral scale, which measures where most magnetic energy
is concentrated and which is given by
\be
l_0=\left.\int (2\pi/k)E_M(k)dk\right/\int E_M(k)dk,
\ee
where $E_M(k)$ is the magnetic energy spectrum,
increases with the Hubble time approximately like $t_H^{0.25}$.

However, around the time of
recombination the photon mean free path $\lambda_\gamma$ became very
large and photon diffusion became very efficient in smoothing out
virtually all inhomogeneities of the photon-baryon plasma \cite{silk}.
This process is often referred to as Silk damping, which corresponds to
a kinematic viscosity $\nu\simeq\lambda_\gamma$ (in natural units).
Silk damping may thus destroy the magnetic field,
as has been noted by Jedamzik, Katalinic and Olinto \cite{chicago}.
Therefore one has to follow numerically the evolution of the magnetic and kinetic
energy spectra in the presence of kinematic viscosity. 
The results \cite{beo2}  are presented in Fig. 5 and the
main point can be summarized as follows: in the cascade models
magnetic energy is transferred to large length scales even in the presence of 
large viscosity. Here the initial magnetic spectrum was chosen to be flat 
in accordance with the large time behaviour suggested by Fig. 4.
\begin{figure}
\leavevmode
\centering
\vspace*{80mm}
\includegraphics{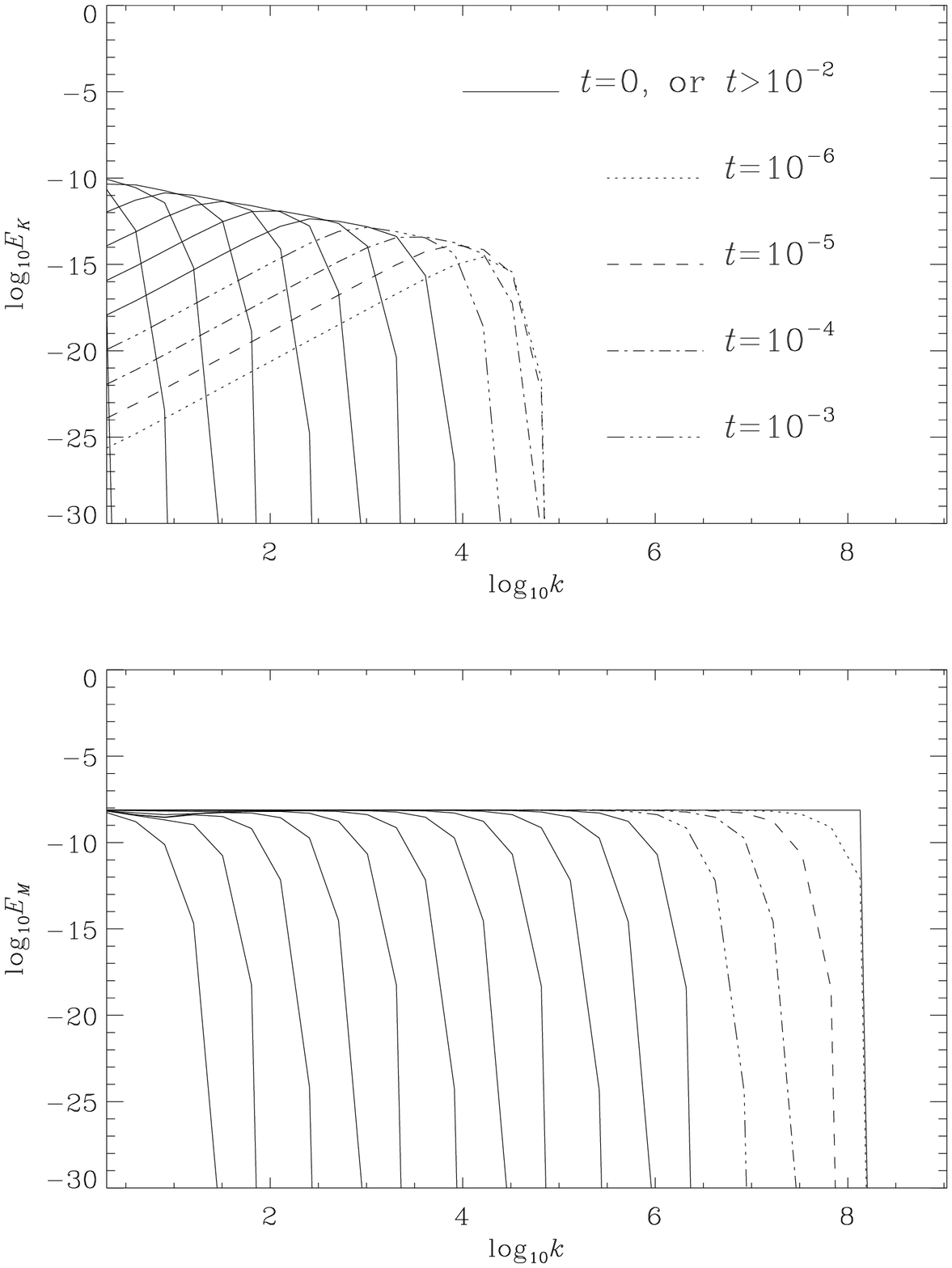}
\includegraphics{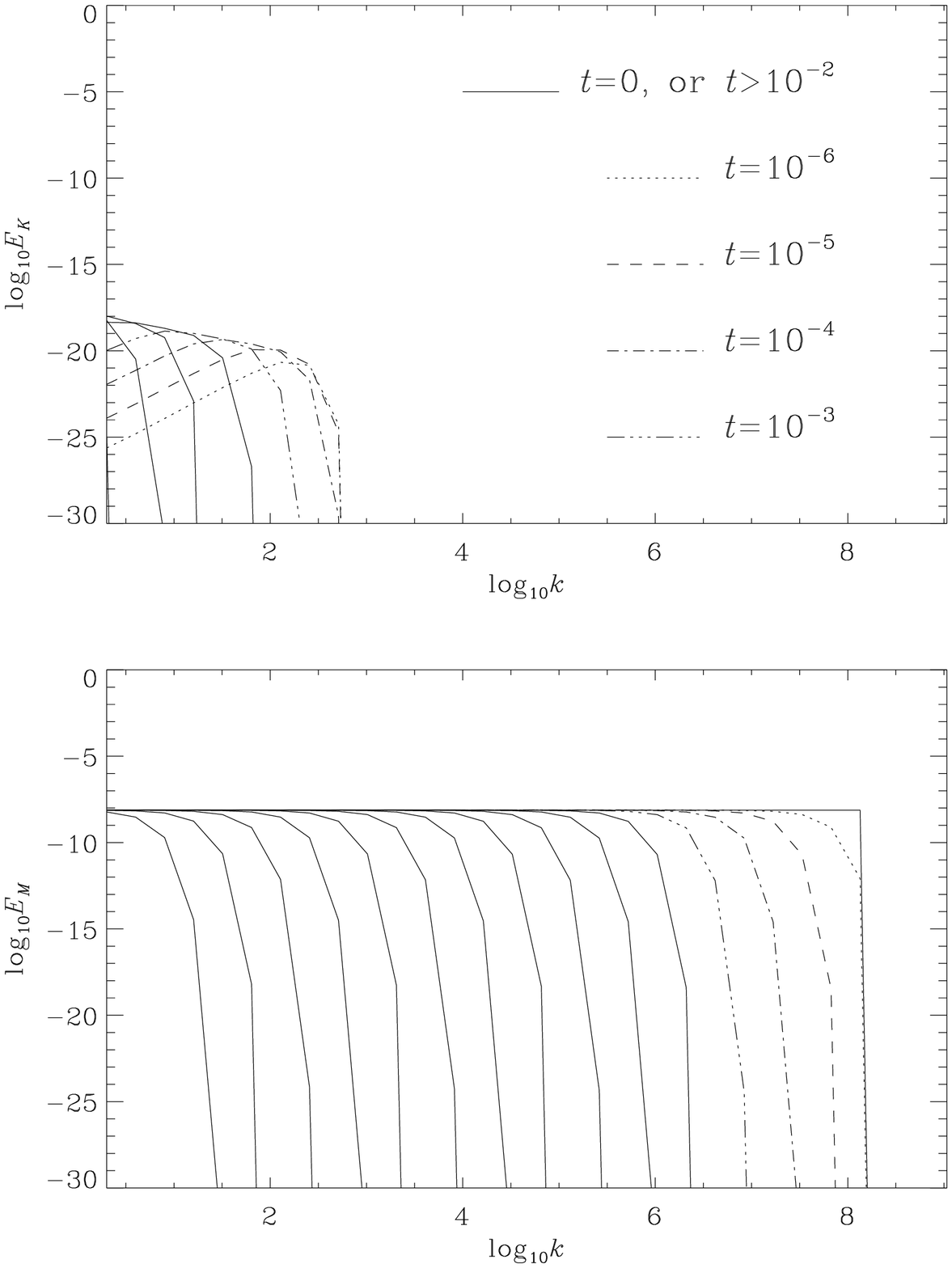}
\caption{Magnetic and plasma kinetic energy spectrum 
as a function of the wave number $k$ in the cascade model for  
small ($\nu=10^{-2}$) plasma viscosity (left) and large ($\nu=10^{2}$) 
plasma viscosity (right). The highest time is $t=10^{8}$, corresponding to
a Hubble time $10^{16}.$}
\label{kuvaaxel}
\end{figure}

For a sufficiently large viscosity, the inverse cascade
stops. One may estimate \cite{beo2} 
that this typically takes place close to recombination.
The results suggest that in the real MHD, inverse cascade is
operative and is essentially not affected by Silk damping, except
very late and perhaps for very weak fields.
Thus we may conclude that it is unlikely that an equipartition
exists in the very early universe. A similar conclusion can be drawn in
a different, continuous  model where the inverse cascade can be found
analytically in an appropriate scaling regime \cite{beo2,poulin}.

\section{Conclusions}
Explaining the galactic magnetic fields in terms of microphysical
processes
that took place when the universe was only ten billionth of a second
old is a daunting task, which is not made easier by the complicated
evolution of the magnetic field as it is twisted and tangled
by the flow of plasma. It is nevertheless encouraging that 
mechanisms for generating primordial magnetic  fields of suitable size exist,
and in particular those based on the early cosmological
phase transitions discussed in Sect. 3 look promising.
At the same time the fact that there are so 
many possibilities tend to underline
our ignorance of the details of the subsequent evolution of the
magnetic field. The step from microphysics to macroscopic fields
is a difficult one because of the very large magnetic Reynolds number
of the early universe. However, different considerations, both analytic
approximations, 2d simulations, as well as the full-fledged shell
model computations which can account for turbulence, seem to point
to the existence of an inverse cascade of magnetic energy. Moreover,
as discussed in Sect. 4.3, the inverse cascade is obtained also in
the presence of a large plasma viscosity. Therefore the primordial
origin of the galactig magnetic fields is quite possible.

Much theoretical work remains to be done, though. At the same time
it is very important that progress is made on the observational
front. In particular, measuring or setting a stringent limit on the
intergalactic field, which could be possible in the near future
as indicated in Sect. 2.3, would provide the testing ground for  
all theoretical scenarios. 

\end{document}